\documentclass[12pt,preprint,notoc,nohyper]{ARXblank} % 10pt is ignored!

\usepackage{epsfig,multicol,bbm}

\newcommand\fverb{\setbox\pippobox=\hbox\bgroup\verb}
\newcommand\fverbdo{\egroup\medskip\noindent%
            \fbox{\unhbox\pippobox}\ }
\newcommand\fverbit{\egroup\item[\fbox{\unhbox\pippobox}]}
\newbox\pippobox
\title{Approximate symmetries of geodesic equations on 2-spheres}

\author{K. Saifullah$^a$ ~and K. Usman$^b$ \\

$^a$Department of Mathematics, Quaid-i-Azam University, Islamabad,
Pakistan \\
$^b$Institute of Applied Mathematics, Technische Universit\"{a}t, Dortmund, Germany \\

Electronic address: \email{saifullah@qau.edu.pk},
\email{kusman@math.uni-dortmund.de}}

\preprint{}  % OR: \preprint{Aaaa/Mm/Yy\\Aaa-aa/Nnnnnn}
\abstract{Approximate symmetries of geodesic equations on 2-spheres
are studied. These are the symmetries of the perturbed geodesic
equations which represent approximate path of a particle rather than
exact path. After giving the exact symmetries of the geodesic
equations, two different approaches to study the approximate
symmetries of the approximate geodesic equations show that no
non-trivial approximate symmetry for these spaces exists.}

\dedicated{}

\begin{document}

\section{Introduction}
For nonlinear problems, analytic solutions are rare and hard to
obtain. Lie group theory provide a systematic and unified approach
in search of analytical solutions \cite{blm, ibr, ptr}. Instead of
solving the nonlinear equations directly, a system of over
determined linear equations is studied using the theory. Once the
so-called symmetries of the equations are calculated, similarity
solutions can be produced. By defining canonical coordinates, it is
also possible to transform the equations into a much simpler form.
Another approach allows to find approximate symmetries of
approximate differential equations. Generally, the perturbed term in
a differential equation corresponds to some small error or
correction. Therefore the resulting equations are defined
approximately depending on a small parameter and such equations
occur frequently in applications. The theory of approximate groups
provides a regular method of calculating the perturbation directly,
without using the complicated group transformations. The calculation
is based on approximate symmetry equations. Recently, approximate
symmetries have been used to find the approximate solutions of some
partial differential equations \cite{C, D} as well.

 Let us consider an approximate (perturbed) equation in variables $t$ and $y$,
 with a small parameter
 $\epsilon$, i.e.
 \begin{equation}
 E(t,y,\epsilon )\approx 0 .  \label{1.55}
 \end{equation}
 We write it as
 \begin{equation}
 E(t,y,\epsilon )\equiv E_{0}(t,y)+\epsilon E_{1}(t,y)\approx 0 ,
 \label{1.56}
 \end{equation}
 where $E_{0}$ and $E_{1}$ are respectively the exact and approximate
 parts of the approximate differential equation (\ref{1.55}). For
 (\ref{1.56}) the infinitesimal generator can be written as \cite{blm, ibr}
 \begin{equation}
 \mathbf{X}=\mathbf{X}_{0}+\epsilon \mathbf{X}_{1},  \label{1.57}
 \end{equation}
 where
 \begin{eqnarray*}
 \mathbf{X}_{0} &=&\xi _{0}\frac{\partial }{\partial t}+\eta
 _{0}\frac{\partial }{\partial y}, \\
 \mathbf{X}_{1} &=&\xi _{1}\frac{\partial }{\partial t}+\eta
 _{1}\frac{\partial }{\partial y}.
 \end{eqnarray*}
 Here $\xi _{0}$, $\xi _{1}$, $\eta _{0}$ and $\eta _{1}$ are all
 functions of $t$ and $y$. \textbf{$X$}$_{0}$ is the generator of the
 exact differential equation $E_{0}$, and \textbf{$X$}$_{1}$ is the
 generator of the approximate differential equation $E_{1}$. Hence we
 can write
 \begin{equation}\nonumber
 \mathbf{X}=\left( \xi _{0}+\epsilon \xi _{1}\right) \frac{\partial }{%
 \partial t}+\left( \eta _{0}+\epsilon \eta _{1}\right) \frac{\partial }{%
 \partial y}.
 \end{equation}
 More generally
 \begin{equation}\nonumber
 \mathbf{X}=\xi \frac{\partial }{\partial t}+\eta \frac{\partial
 }{\partial y} ,
 \end{equation}
 where
 \begin{eqnarray*}
 \xi &=&\xi _{0}+\epsilon \xi _{1}, \\
 \eta &=&\eta _{0}+\epsilon \eta _{1}.
 \end{eqnarray*}
 A second prolongation \cite{blm, ibr} is given by
 \begin{equation}\nonumber
 \mathbf{X}=\xi \frac{\partial }{\partial t}+\eta \frac{\partial }{\partial
 y}+\eta ^{(1)}\frac{\partial }{\partial y^{\prime }}+\eta ^{(2)}
 \frac{\partial }{\partial y^{\prime \prime }} .
 \end{equation}
 Equation (\ref{1.56}) is approximately invariant under the
 approximate group of transformations with the generator given in
 (\ref{1.57}) if and only if $\mathbf{X}E|_{E\approx
 0}=o(\epsilon )$ \cite{ibr}, or
 \begin{equation}
 \lbrack \mathbf{X}_{0}E_{0}+\epsilon (\mathbf{X}_{0}E_{1}+\mathbf{X}
 _{1}E_{0})]{\LARGE \mid }_{E\approx 0}=o(\epsilon ) .
 \label{1.58}
 \end{equation}
 This is the \textit{determining equation} and \textbf{$X$} is called an
 \textit{infinitesimal approximate symmetry} or an \textit{approximate symmetry}
 or an \textit{approximate operator} \cite{ibr},
 admitted by (\ref{1.56}) .

 If \textbf{$X$}$_{0}$ is a generator of the unperturbed differential
 equation
 \begin{equation}
 E_{0}=0 ,  \label{1.59}
 \end{equation}
 obtained by solving the determining equation
 \begin{equation}
 \mathbf{X}_{0}E_{0}{\LARGE \mid }_{E\approx 0}=0 ,
 \label{1.60}
 \end{equation}
 then we define the auxiliary function $H$ by
 \begin{equation}
 H=\frac{1}{\epsilon }\mathbf{X}_{0}(E_{0}+\epsilon E_{1}){\LARGE \mid }%
 _{E_{0}+\epsilon E_{1}=0} ,  \label{1.61}
 \end{equation}
 and find an \textit{approximate symmetry}, (\ref{1.57}), of the
 perturbed differential equation (\ref{1.56}) by solving for
 \textbf{$X$}$_{1}$\cite{ibr} in
 \begin{equation}
 \mathbf{X}_{1}E_{0}{\LARGE \mid }_{E\approx 0}+H=0 . \label{1.62}
 \end{equation}

 Note that (\ref{1.62}), unlike the determining equation (\ref{1.60})
 for exact symmetries, is non-homogeneous.

 The geodesic equations on a manifold can be regarded as a system of
 second order ordinary differential equations. The curvature of
 surfaces can be positive, zero or negative. Symmetries of geodesic equations for these
 three types of manifolds have been studied \cite{A}. Interesting
 connection between these symmetries and the isometries \cite{oni} of
 these spaces have also been found. In this paper we explore
 approximate symmetries of geodesic equations of 2-spheres. Perturbations in geodesic equations correspond
 to approximate path of the particle as opposed to the exact one. It
 may be noted that we study approximate geodesics on a 2-manifold
 rather than exact geodesics on a perturbed manifold.

 \section{Approximate symmetries of geodesic equations on a 2-sphere}

 A geodesic is a locally length-minimizing curve and give the
 shortest distance between two points. Equivalently, it is a
 path that a particle which is not accelerating would follow. The
 geodesic equations on a manifold can be regarded as a system of
 second order ordinary differential equations. In coordinates $x^{a}$
 these can be expressed as \cite{onl, miln}
 \begin{equation}
 \ddot{x}^{a}+\Gamma _{bc}^{a}\dot{x}^{b}\dot {x}^{c}=0,
 \label{a}
 \end{equation}
 where dot represents derivative with respect to the arc length
 parameter $s$ and $\Gamma _{bc}^{a}$ denote the Christoffel symbols,
 which for a metric $g_{ab}$, are defined by
 \begin{equation}
 \Gamma _{bc}^{a}=\frac{1}{2}g^{ad}\left( g_{bc,d}+g_{bd,c}-g_{cd,b}\right) .
 \end{equation}
Here $g^{ab}$ denotes the inverse of the metric and the range of the
indices gives the dimension of the space.

 Exact symmetries of the unit sphere have been given in Ref. \cite{A}. We first
 outline  the derivation of these symmetries and then use them for
 constructing the approximate symmetries. The metric for the
 sphere can be written as
 \begin{equation}\nonumber
 ds^{2}=d\theta ^{2}+\sin ^{2}\theta d\phi ^{2}.
 \end{equation}
The geodesic equations (\ref{a}) for this metric are given by
 \begin{eqnarray}
 \nonumber E_{1} &:& \ddot{\theta }-\sin \theta \cos \theta \dot{
 \phi }^{2}=0,  \\
 E_{2} &:& \ddot{\phi }+2\cot \theta \dot{\theta}
 \dot{\phi }=0.  \label{2}
 \end{eqnarray}
 These are second order ordinary differential equations. We apply the second
 prolongation \cite{A},
 \begin{equation}
 \mathbf{X}=\xi \frac{\partial }{\partial s}+\eta ^{1}\frac{\partial }{%
 \partial \theta }+\eta ^{2}\frac{\partial }{\partial \phi }+\eta _{,s}^{1}%
 \frac{\partial }{\partial \dot{\theta }}+\eta _{,s}^{2}\frac{\partial
 }{\partial \dot{\phi }}+\eta _{,ss}^{1}\frac{\partial }{\partial
 \ddot{\theta }}+\eta _{,ss}^{2}\frac{\partial }{\partial \ddot{%
 \phi }},  \label{3}
 \end{equation}
 of the symmetry generator given by (\ref
 {1.57}), to both the geodesic equations. Here $\xi $, $\eta ^{1}$ and $\eta ^{2}$
 are all functions of $s$, $\theta $
 and $\phi $; $\eta _{,s}^{1}$ and $\eta _{,s}^{2}$ are functions of $s$, $%
 \theta $, $\phi $, $\dot{\theta }$ and $\dot{\phi }$; and $%
 \eta _{,ss}^{1}$ and $\eta _{,ss}^{2}$ are all functions of $s$, $\theta $, $%
 \phi $, $\dot{\theta }$, $\dot{\phi }$, $\ddot{\theta }$
 and $\ddot{\phi }$, Then
 \begin{eqnarray*}
 \mathbf{X}E_{1} &{\LARGE \mid }&_{E_{1}=0=E_{2}}=0, \\
 \mathbf{X}E_{2} &{\LARGE \mid }&_{E_{1}=0=E_{2}}=0,
 \end{eqnarray*}
 respectively yield \cite{A}
 \begin{equation}
 \left[ \eta _{,ss}^{1}-2\sin \theta \cos \theta \dot{\phi
 }\eta
 _{,s}^{2}-\eta ^{1}\left( \cos ^{2}\theta -\sin ^{2}\theta \right)
 \dot{\phi }^{2}\right] {\LARGE \mid }_{E_{1}=0=E_{2}}=0,  \label{4}
 \end{equation}
 \begin{equation}
 \left[ \eta _{,ss}^{2}+2\cot \theta \left( \dot{\phi }\eta _{,s}^{1}+%
 \dot{\theta }\eta _{,s}^{2}\right) -2\dot{\theta }\dot{
 \phi }\csc ^{2}\theta \eta ^{1}\right] {\LARGE \mid
 }_{E_{1}=0=E_{2}}=0. \label{5}
 \end{equation}
 If we write
 \begin{equation}
 D=\frac{\partial }{\partial s}+\dot{\theta }\frac{\partial
 }{\partial
 \theta }+\dot{\phi }\frac{\partial }{\partial \phi }+\ddot{%
 \theta }\frac{\partial }{\partial \dot{\theta }}+\ddot{\phi }%
 \frac{\partial }{\partial \dot{\phi }},  \label{6}
 \end{equation}
 then

 \begin{equation}
 \eta _{,s}^{1}=D\eta ^{1}-\dot{\theta }D\xi =\eta _{s}^{1}+\dot
 {\theta }\left( \eta _{\theta }^{1}-\xi _{s}\right) -\dot{\theta }%
 ^{2}\xi _{\theta }+\dot{\phi }\eta _{\phi }^{1}-\dot{\theta }%
 \dot{\phi }\xi _{\phi },  \label{7}
 \end{equation}
 \begin{equation}
 \eta _{,s}^{2}=D\eta ^{2}-\dot{\phi }D\xi =\eta _{s}^{2}+\dot{%
 \phi }\left( \eta _{\phi }^{2}-\xi _{s}\right) +\dot{\theta
 }\eta
 _{\theta }^{2}-\dot{\phi ^{2}}\xi _{\phi }-\dot
 {\theta }%
 \dot{\phi }\xi _{\theta },  \label{8}
 \end{equation}
 \begin{eqnarray}
 \nonumber \eta _{,ss}^{1} &=&D\eta _{,s}^{1}-\ddot{\theta }D\xi    \\
 \nonumber &=&\eta _{ss}^{1}+\dot{\theta }\left( 2\eta _{s\theta
 }^{1}-\xi _{ss}\right) +\dot{\theta }^{2}\left( \eta _{\theta
 \theta }^{1}-2\xi
 _{s\theta }\right) -\dot{\theta }^{3}\xi _{\theta \theta }+2\dot {\phi }\eta _{s\phi }^{1}  \\
 \nonumber &&+\dot{\phi }^{2}\eta _{\phi \phi }^{1}+2\dot{\theta }
 \dot{\phi }\left( \eta _{\theta \phi }^{1}-\xi _{s\phi }\right) -2\dot
 {%
 \theta }^{2}\dot{\phi }\xi _{\theta \phi }-\dot{\theta }%
 \dot{\phi }^{2}\xi _{\phi \phi }  \\
 &&+\ddot{\theta }\left( \eta _{\theta }^{1}-2\xi _{s}-3\dot{%
 \theta }\xi _{\theta }\right) -2\ddot{\theta
 }\dot{\phi }\xi
 _{\phi }+\ddot{\phi }\eta _{\phi }^{1}-\dot{\theta }\ddot{\phi }\xi _{\phi },  \label{9}
 \end{eqnarray}
 \begin{eqnarray}
 \nonumber \eta _{,ss}^{2} &=&D\eta _{,s}^{2}-\ddot{\phi }D\xi   \\
 \nonumber &=&\eta _{ss}^{2}+2\dot{\theta }\eta _{s\theta }^{2}+\dot{%
 \theta }^{2}\eta _{\theta \theta }^{2}+\dot{\phi }\left(
 2\eta _{s\phi }^{2}-\xi _{ss}\right) +\dot{\phi }^{2}\left(
 \eta _{\phi
 \phi }^{2}-2\xi _{s\phi }\right)  \\
 \nonumber &&-\dot{\phi }^{3}\xi _{\phi \phi }+2\dot{\theta }\dot{%
 \phi }\left( \eta _{\theta \phi }^{2}-\xi _{s\theta }\right) -\dot{%
 \theta }^{2}\dot{\phi }\xi _{\theta \theta }-2\dot{\theta }%
 \dot{\phi }^{2}\xi _{\theta \phi }+\ddot{\theta }\eta
 _{\theta
 }^{2}  \\
 &&-\ddot{\theta }\dot{\phi }\xi _{\theta }+\ddot{\phi }%
 \left( \eta _{\phi }^{2}-2\xi _{s}-3\dot{\phi }\xi _{\phi }\right) -2
 \dot{\theta }\ddot{\phi }\xi _{\theta }.  \label{10}
 \end{eqnarray}
 Inserting (\ref{7}) - (\ref{10}) into (\ref{4}) and (\ref{5}) yields
 \begin{eqnarray}
 \nonumber \lbrack \eta _{ss}^{1}+\dot{\theta }\left( 2\eta _{s\theta
 }^{1}-\xi _{ss}\right) +\dot{\theta }^{2}\left( \eta _{\theta
 \theta }^{1}-2\xi
 _{s\theta }\right) -\dot{\theta }^{3}\xi _{\theta \theta }+2\dot{\phi }\eta _{s\phi }^{1}+\dot{\phi }^{2}\eta _{\phi \phi
 }^{1}
 \\
 \nonumber +2\dot{\theta }\dot{\phi }\left( \eta _{\theta \phi
 }^{1}-\xi _{s\phi }\right) -2\dot{\theta
 }^{2}\dot{\phi }\xi _{\theta
 \phi }-\dot{\theta }\dot{\phi }^{2}\xi _{\phi \phi }+\ddot{\theta }\left( \eta _{\theta }^{1}-2\xi _{s}-3\dot{\theta
 }\xi
 _{\theta }\right)  \\
 -2\ddot{\theta }\dot{\phi }\xi _{\phi }+\ddot{\phi }%
 \eta _{\phi }^{1}-\dot{\theta }\ddot{\phi }\xi _{\phi
 }]-\sin 2\theta \dot{\phi }[\eta _{s}^{2}+\dot{\phi
 }\left( \eta
 _{\phi }^{2}-\xi _{s}\right) +\dot{\theta }\eta _{\theta }^{2}-%
 \dot{\phi ^{2}}\xi _{\phi }  \nonumber \\
 -\dot{\theta }\dot{\phi }\xi _{\theta }]-\eta ^{1}\cos
 2\theta \dot{\phi }^{2}|_{E_{1}=0=E_{2}}=0,  \label{11}
 \end{eqnarray}
 \begin{eqnarray}
 \nonumber \lbrack \eta _{ss}^{2}+2\dot{\theta }\eta _{s\theta }^{2}+\dot{
 \theta }^{2}\eta _{\theta \theta }^{2}+\dot{\phi }\left(
 2\eta _{s\phi }^{2}-\xi _{ss}\right) +\dot{\phi }^{2}\left(
 \eta _{\phi
 \phi }^{2}-2\xi _{s\phi }\right) -\dot{\phi }^{3}\xi _{\phi \phi } \\
 \nonumber +2\dot{\theta }\dot{\phi }\left( \eta _{\theta \phi
 }^{2}-\xi
 _{s\theta }\right)
 -\dot{\theta }^{2}\dot{\phi }\xi _{\theta \theta }-2\dot
 {\theta }\dot{\phi }^{2}\xi _{\theta \phi
 }+\ddot{\theta }\eta
 _{\theta }^{2}-\ddot{\theta }\dot{\phi }\xi _{\theta }+\ddot{\phi }\left( \eta _{\phi }^{2}-2\xi _{s}-3\dot{\phi }\xi
 _{\phi }\right) \\
 \nonumber -2\dot{\theta }\ddot{\phi }\xi
 _{\theta }]
 +2\cot \theta
 (\dot{\phi }\{\eta _{s}^{1}+\dot{\theta }\left( \eta
 _{\theta
 }^{1}-\xi _{s}\right) -\dot{\theta }^{2}\xi _{\theta }+\dot{
 \phi }\eta _{\phi }^{1}-\dot{\theta }\dot{\phi }\xi _{\phi
 }\}  \\
 + \dot{\theta }\{\eta _{s}^{2}+\dot{\phi }\left( \eta
 _{\phi
 }^{2}-\xi _{s}\right)
 +\dot{\theta }\eta _{\theta }^{2}-\dot{\phi ^{2}}\xi _{\phi }-%
 \dot{\theta }\dot{\phi }\xi _{\theta }\})-2\dot{\theta }%
 \dot{\phi }\csc ^{2}\theta \eta ^{1}|_{E_{1}=0=E_{2}}=0.
 \label{12}
 \end{eqnarray}
 Inserting the values of $\ddot{\theta }$ and
 $\ddot{\phi }$ from the geodesic equations (\ref{2}), and
 then comparing the
 coefficients of the powers of $\dot{\theta }$ and $\dot{\phi }$,
 we obtain the following system of partial differential equations.
 \begin{eqnarray}
 (\dot{\theta }\dot{\phi })^{0} &:&\eta _{ss}^{1}=0,
 \label{13a} \\
 \eta _{ss}^{2} &=&0,  \label{13b}
 \end{eqnarray}
 \begin{eqnarray}
 \dot{\theta }:2\eta _{s\theta }^{1}-\xi _{ss}=0,  \label{14a} \\
 \eta _{s\theta }^{2}+\cot \theta \eta _{s}^{2}=0,  \label{14b}
 \end{eqnarray}
 \begin{eqnarray}
 \dot{\phi }:2\eta _{s\phi }^{1}-\sin 2\theta \eta _{s}^{2}=0,
 \label{15a} \\
 2\eta _{s\phi }^{2}-\xi _{ss}+2\cot \theta \eta _{s}^{1}=0,
 \label{15b}
 \end{eqnarray}
 \begin{eqnarray}
 \dot{\theta }^{2}:\eta _{\theta \theta }^{1}-2\xi _{s\theta
 }=0,
 \label{16a} \\
 \eta _{\theta \theta }^{2}+2\cot \theta \eta _{\theta }^{2}=0,
 \label{16b}
 \end{eqnarray}
 \begin{eqnarray}
 \dot{\phi }^{2}:\eta _{\phi \phi }^{1}+\sin \theta \cos
 \theta \eta _{\theta }^{1}-\eta ^{1}\cos 2\theta -\sin 2\theta \eta
 _{\phi }^{2}=0,
 \label{17a} \\
 \eta _{\phi \phi }^{2}-2\xi _{s\phi }+\sin \theta \cos \theta \eta
 _{\theta }^{2}+2\cot \theta \eta _{\phi }^{1}=0,  \label{17b}
 \end{eqnarray}
 \begin{equation}
 \dot{\theta }^{3}:\xi _{\theta \theta }=0,  \label{18}
 \end{equation}
 \begin{equation}
 \dot{\phi }^{3}:-\xi _{\phi \phi }-\sin \theta \cos \theta
 \xi _{\theta }=0,  \label{19}
 \end{equation}
 \begin{eqnarray}
 \dot{\theta }\dot{\phi }:2\left( \eta _{\theta \phi
 }^{1}-\xi _{s\phi }\right) -\sin 2\theta \eta _{\theta }^{2}-2\cot
 \theta \eta _{\phi
 }^{1}=0,  \label{20a} \\
 \left( \eta _{\theta \phi }^{2}-\xi _{s\theta }\right) +\cot
 \theta \eta _{\theta }^{1}-\csc ^{2}\theta \eta ^{1}=0,
 \label{20b}
 \end{eqnarray}
 \begin{eqnarray}
 \dot{\theta }\dot{\phi }^{2}:-\xi _{\phi \phi }-\sin
 \theta \cos \theta
 \xi _{\theta }=0,  \label{21a} \\
 -2\xi _{\theta \phi }+2\cot \theta \xi _{\phi }=0 .  \label{21b}
 \end{eqnarray}
 Equations (\ref{18}), (\ref{19}) and (\ref{21a}) yield
 \begin{equation}
 \xi =b_{1}{\small (}s{\small )}.  \label{22}
 \end{equation}
 Substituting the value of $\xi $ into (\ref{16a}) and solving with
 (\ref{13a}), we get
 \begin{equation}
 \eta ^{1}=\left[ f_{1}{\small (}\phi {\small )}s+f_{2}{\small (}\phi {\small %
 )}\right] \theta +f_{3}{\small (}\phi {\small )}s+f_{4}{\small
 (}\phi {\small )}.  \label{23}
 \end{equation}
 From (\ref{13b}), (\ref{14b}) and (\ref{16b}) we obtain
 \begin{equation}
 \eta ^{2}=-\cot \theta f_{6}{\small (}\phi {\small )}+f_{7}{\small
 (}\phi {\small )}.  \label{24}
 \end{equation}
 Substituting the values of $\eta ^{1}$ and $\eta ^{2}$ into (\ref{15a}) we get
 \begin{equation}
 \eta ^{1}=\left[ As+f_{2}{\small (}\phi {\small )}\right] \theta +Bs+f_{4}%
 {\small (}\phi {\small )}.  \label{25}
 \end{equation}
 Then (\ref{14a}) yields
 \begin{equation}
 \xi =As^{2}+c_{1}s+c_{0}.  \label{26}
 \end{equation}
 Now (\ref{17a}) and (\ref{20a}) give
 \begin{equation}
 \xi =c_{1}s+c_{0},  \label{27}
 \end{equation}
 \begin{equation}
 \eta ^{1}=c_{3}\cos \phi +c_{4}\sin \phi ,  \label{28}
 \end{equation}
 \begin{equation}
 \eta ^{2}=\cot \theta \left( c_{4}\cos \phi -c_{3}\sin \phi \right)
 +c_{2}. \label{29}
 \end{equation}
 Therefore, the exact symmetries come out to be
 \begin{eqnarray}
 \mathbf{X}_{0} &=&\frac{\partial }{\partial s},\mathbf{X}_{1}=s%
 \frac{\partial }{\partial s},\mathbf{X}_{2}=\frac{\partial }{%
 \partial \phi }, \nonumber \\
 \mathbf{X}_{3} &=& \cos \phi \frac{\partial }{\partial
 \theta }-\cot \theta \sin \phi \frac{\partial }{\partial \phi },  \nonumber \\
 \mathbf{X}_{4} &=&\sin \phi \frac{\partial }{\partial \theta }+\cot
 \theta \cos \phi \frac{\partial }{\partial \phi }.  \label{30}
 \end{eqnarray}

 In the next sections we discuss two different approaches for finding
 approximate symmetries of this 2-manifold.

 \subsection{Approximate symmetries: First approach}

 Here we will convert the geodesic equations (\ref{2}) of the sphere
 into perturbed equations by adding a general function and then try
 to find the approximate symmetries of these equations.

 The second prolongations of the five exact symmetries (\ref{30}) of
 the geodesic equations of the sphere are given below.

 (i)$\qquad \mathbf{X}^{0}=\frac{\partial }{\partial s}$

 Here $\xi =1,$ $\eta ^{1}=0=\eta ^{2},$ therefore, equations (\ref{7})
 -(\ref{10}) give $\eta _{,s}^{1}=0=\eta _{,s}^{2}$,
 $\eta_{,ss}^{1}=0=\eta _{,ss}^{2}$ so that the second prolongation
 becomes
 \begin{equation}\nonumber
 \mathbf{X}^{0}=\frac{\partial }{\partial s}.
 \end{equation}

 (ii)$\qquad \mathbf{X}^{1}=s\frac{\partial }{\partial s}$

 Here $\xi =s,$ $\eta ^{1}=0=\eta ^{2}$, therefore, equations
 (\ref{7}) -(\ref{10}) give $\eta _{,s}^{1}=-\dot{\theta }$,
 $\eta _{,s}^{2}=-\dot{\phi }$, $\eta
 _{,ss}^{1}=-2\ddot{\theta },$ $\eta
 _{,ss}^{2}=-2\ddot{\phi },$
 so that the second prolongation becomes
 \begin{equation}
 \mathbf{X}^{1}=s\frac{\partial }{\partial s}-\dot{\theta }\frac{%
 \partial }{\partial \dot{\theta }}-\dot{\phi }\frac{\partial }{%
 \partial \dot{\phi }}-2\ddot{\theta }\frac{\partial }{\partial
 \ddot{\theta }}-2\ddot{\phi }\frac{\partial }{\partial \ddot{\phi }}.  \label{32}
 \end{equation}

 (iii)$\qquad \mathbf{X}^{2}=\frac{\partial }{\partial \phi }$

 Here $\xi =0,$ $\eta ^{1}=0,$ $\eta ^{2}=1$, therefore, equations
 (\ref{7}) -(\ref{10}) give $\eta _{,s}^{1}=0=$ $\eta _{,s}^{2}$,
 $\eta _{,ss}^{1}=0=$ $\eta _{,ss}^{2}$ so that second prolongation
 becomes
 \begin{equation}
 \mathbf{X}^{2}=\frac{\partial }{\partial \phi }.  \label{33}
 \end{equation}

 (iv)$\qquad \mathbf{X}^{3}=\cos \phi \frac{\partial }{\partial \theta }-\cot
 \theta \sin \phi \frac{\partial }{\partial \phi }0.$

 Here $\xi =0,$ $\eta ^{1}=\cos \phi ,$ $\eta ^{2}=-\cot \theta \sin
 \phi $, therefore, equations (\ref{7}) -(\ref{10}) give
 \begin{eqnarray*}
 \eta _{,s}^{1} &=&-\dot{\phi }\sin \phi , \\
 \eta _{,s}^{2} &=&\dot{\theta }\left( \sin \phi \csc ^{2}\theta
 \right) +\dot{\phi }\left( -\cot \theta \cos \phi \right) , \\
 \eta _{,ss}^{1} &=&-\dot{\phi }^{2}\cos \phi -\ddot{\phi }\sin
 \phi , \\
 \eta _{,ss}^{2} &=&\dot{\theta }^{2}\sin \phi \left( -2\csc
 ^{2}\theta \cot \theta \right) +\dot{\phi }^{2}\cot \theta \sin \phi
 +2\dot{\theta }\dot{\phi }\left( \cos \phi \csc ^{2}\theta
 \right) \\
 &&+\ddot{\theta }\csc ^{2}\theta \sin \phi +\ddot{\phi }\left(
 -\cot \theta \cos \phi \right) ,
 \end{eqnarray*}
 so that the second prolongation becomes
 \begin{eqnarray}
 \mathbf{X}^{3} &=&\cos \phi \frac{\partial }{\partial \theta }-\cot \theta
 \sin \phi \frac{\partial }{\partial \phi }-\dot{\phi }\sin \phi \frac{%
 \partial }{\partial \dot{\theta }}+\left( \dot{\theta }\sin
 \phi \csc ^{2}\theta -\dot{\phi }\cot \theta \cos \phi \right) \frac{%
 \partial }{\partial \dot{\phi }}  \nonumber \\
 &&-\left( \dot{\phi }^{2}\cos \phi +\ddot{\phi }\sin \phi
 \right) \frac{\partial }{\partial \ddot{\theta }}+(-2\dot{
 \theta }^{2}\sin \phi \csc ^{2}\theta \cot \theta +\dot{\phi}
 ^{2}\cot \theta \sin \phi  \nonumber \\
 &&+2\dot{\theta }\dot{\phi }\cos \phi \csc ^{2}\theta +\ddot{\theta }\csc ^{2}\theta \sin \phi -\ddot{\phi }\cot \theta \cos
 \phi )\frac{\partial }{\partial \ddot{\phi }}.  \label{34}
 \end{eqnarray}

 (v)$\qquad \mathbf{X}^{4}=\sin \phi \frac{\partial }{\partial \theta }+\cot
 \theta \cos \phi \frac{\partial }{\partial \phi }$

 Here $\xi =0,$ $\eta ^{1}=\sin \phi ,$ $\eta ^{2}=\cot \theta \cos
 \phi $, therefore, equations (\ref{7}) -(\ref{10}) give
 \begin{eqnarray*}
 \eta _{,s}^{1} &=&\dot{\phi }\cos \phi , \\
 \eta _{,s}^{2} &=&-\dot{\theta }\csc ^{2}\theta \cos \phi -\dot%
 {\phi }\cot \theta \sin \phi , \\
 \eta _{,ss}^{1} &=&-\dot{\phi }^{2}\sin \phi +\ddot{\phi }\cos
 \phi , \\
 \eta _{,ss}^{2} &=&\dot{\theta }^{2}\left( 2\csc ^{2}\theta \cot
 \theta \cos \phi \right) +\dot{\phi }^{2}\left( -\cot \theta \cos
 \phi \right) +2\dot{\theta }\dot{\phi }\left( \csc ^{2}\theta
 \sin \phi \right) \\
 &&-\ddot{\theta }\csc ^{2}\theta \cos \phi -\ddot{\phi }\cot
 \theta \sin \phi ,
 \end{eqnarray*}
 so that the second prolongation becomes
 \begin{eqnarray}
 \mathbf{X}^{4} &=&\sin \phi \frac{\partial }{\partial \theta }+\cot \theta
 \cos \phi \frac{\partial }{\partial \phi }+\dot{\phi }\cos \phi \frac{%
 \partial }{\partial \dot{\theta }}-\left( \dot{\theta }\csc
 ^{2}\theta \cos \phi +\dot{\phi }\cot \theta \sin \phi \right) \frac{%
 \partial }{\partial \dot{\phi }}  \nonumber \\
 &&+\left( \ddot{\phi }\cos \phi -\dot{\phi }^{2}\sin \phi
 \right) \frac{\partial }{\partial \ddot{\theta }}+(2\dot{%
 \theta }^{2}\csc ^{2}\theta \cot \theta \cos \phi +\dot{-\phi }%
 ^{2}\cot \theta \cos \phi  \nonumber \\
 &&+2\dot{\theta }\dot{\phi }\csc ^{2}\theta \sin \phi -\ddot{\theta }\csc ^{2}\theta \cos \phi -\ddot{\phi }\cot \theta \sin
 \phi)\frac{\partial }{\partial \ddot{\phi }}.  \label{35}
 \end{eqnarray}

 Now, we approximate the geodesic equations of a sphere by using two
 arbitrary functions  $f(\theta)$, $g(\phi)$ and a small parameter
 $\epsilon$ as
 \begin{eqnarray}
 E_{1}^{1} &:& \ddot{\theta }-\sin \theta \cos \theta \dot{\phi }^{2}+\epsilon f{\small (}\theta {\small )}=0,  \nonumber \\
 E_{1}^{2} &:& \ddot{\phi }+2\cot \theta \dot{\theta }%
 \dot{\phi }+\epsilon g\left( \phi \right) =0,  \label{36}
 \end{eqnarray}
 where
 \begin{eqnarray}
 E_{0}^{1} &:& \ddot{\theta }-\sin \theta \cos \theta \dot{\phi }^{2}=0,  \nonumber \\
 E_{0}^{2} &:& \ddot{\phi }+2\cot \theta \dot{\theta }%
 \dot{\phi }=0,  \label{36a}
 \end{eqnarray}
 are the exact geodesic equations. For Case (i) the auxiliary functions
 become
 \begin{eqnarray}
 H_{1} &=&\frac{1}{\epsilon }\mathbf{X}_{0}^{0}(\ddot{\theta }-\sin
 \theta \cos \theta \dot{\phi }^{2}+\epsilon f{\small (}\theta {\small %
 )}){\LARGE \mid }_{\ddot{\theta }-\sin \theta \cos \theta \dot{%
 \phi}^{2}+\epsilon f{\small (}\theta {\small )}=0}  \nonumber \\
 &=&\frac{1}{\epsilon }\left( \frac{\partial }{\partial s}\right)
 (\ddot{\theta }-\sin \theta \cos \theta \dot{\phi }^{2}+\epsilon f{\small (
 }\theta {\small)}{\LARGE \mid }_{\ddot{\theta }-\sin \theta \cos
 \theta \dot{\phi }^{2}+\epsilon f{\small (}\theta {\small )}=0}
 \nonumber \\
 &=&\frac{1}{\epsilon }\left( 0\right) =0,  \nonumber
 \end{eqnarray}
 \begin{eqnarray}
 H_{2} &=&\frac{1}{\epsilon }\mathbf{X}_{0}^{0}\left( \ddot{\phi }%
 +2\cot \theta \dot{\theta }\dot{\phi }+\epsilon g{\small (}%
 \phi {\small )}\right) {\LARGE \mid }_{\ddot{\phi }+2\cot \theta
 \dot{\theta }\dot{\phi }+\epsilon g{\small (}\phi {\small )}=0}
 \nonumber \\
 &=&\frac{1}{\epsilon }\left( \frac{\partial }{\partial s}\right) \left(
 \ddot{\phi }+2\cot \theta \dot{\theta }\dot{\phi }%
 +\epsilon g{\small (}\phi {\small )}\right) {\LARGE \mid }_{\ddot{%
 \phi }+2\cot \theta \dot{\theta }\dot{\phi }+\epsilon g{\small %
 (}\phi {\small )}=0}  \nonumber \\
 &=&\frac{1}{\epsilon }\left( 0\right) =0.
 \end{eqnarray}
 Hence $H_{1}=0=H_{2}$, so we cannot proceed further.

 Now we take Case (ii). Here the auxiliary functions become
 \begin{eqnarray}
 H_{1} &=&\frac{1}{\epsilon }\mathbf{X}_{0}^{1}(\ddot{\theta }-\sin
 \theta \cos \theta \dot{\phi }^{2}+\epsilon f{\small (}\theta {\small %
 )}){\LARGE \mid }_{\ddot{\theta }-\sin \theta \cos \theta \dot{%
 \phi }^{2}+\epsilon f{\small (}\theta {\small )}=0}  \nonumber \\
 &=&\frac{1}{\epsilon }\left( s\frac{\partial }{\partial s}-\dot{%
 \theta }\frac{\partial }{\partial \dot{\theta }}-\dot{\phi }%
 \frac{\partial }{\partial \dot{\phi }}-2\ddot{\theta }\frac{%
 \partial }{\partial \ddot{\theta }}-2\ddot{\phi }\frac{%
 \partial }{\partial \ddot{\phi }}\right) \nonumber \\
 &\times&\left(\ddot{\theta }-\sin
 \theta \cos \theta \dot{\phi }^{2}+\epsilon f{\small (}\theta {\small %
 )}\right){\LARGE \mid }_{\ddot{\theta }-\sin \theta \cos \theta \dot{%
 \phi }^{2}+\epsilon f{\small (}\theta {\small )}=0}  \nonumber \\
 &=&\frac{1}{\epsilon }\left( -2\ddot{\theta }+2\sin \theta \cos
 \theta \dot{\phi }^{2}+\epsilon \left( 0\right) \right) {\LARGE \mid }%
 _{\ddot{\theta }-\sin \theta \cos \theta \dot{\phi }%
 ^{2}+\epsilon f{\small (}\theta {\small )}=0} \nonumber \\
 &=& \frac{1}{\epsilon }\left(
 -2\left( -\epsilon f{\small (}\theta {\small )}\right) \right)
 =2f{\small (}\theta {\small )},  \nonumber
 \end{eqnarray}
 \begin{eqnarray}
 H_{2} &=&\frac{1}{\epsilon }\mathbf{X}_{0}^{1}\left( \ddot{\phi }%
 +2\cot \theta \dot{\theta }\dot{\phi }+\epsilon g{\small (}%
 \phi {\small )}\right) {\LARGE \mid }_{\ddot{\phi }+2\cot \theta
 \dot{\theta }\dot{\phi }+\epsilon g{\small (}\phi {\small )}=0}
 \nonumber \\
 &=&\frac{1}{\epsilon }\left( s\frac{\partial }{\partial s}-\dot{%
 \theta }\frac{\partial }{\partial \dot{\theta }}-\dot{\phi }%
 \frac{\partial }{\partial \dot{\phi }}-2\ddot{\theta }\frac{%
 \partial }{\partial \ddot{\theta }}-2\ddot{\phi }\frac{%
 \partial }{\partial \ddot{\phi }}\right) \nonumber \\
 &\times&\left( \ddot{\phi }%
 +2\cot \theta \dot{\theta }\dot{\phi }+\epsilon g{\small (}%
 \phi {\small )}\right) {\LARGE \mid }_{\ddot{\phi }+2\cot \theta
 \dot{\theta }\dot{\phi }+\epsilon g{\small (}\phi {\small )}=0}
 \nonumber \\
 &=&\frac{1}{\epsilon }\left( -2\ddot{\phi }-2\cot \theta \dot{%
 \theta }\dot{\phi }-2\cot \theta \dot{\theta }\dot{\phi
 }+\epsilon \left( 0\right) \right) {\LARGE \mid }_{\ddot{\phi }+2\cot
 \theta \dot{\theta }\dot{\phi }+\epsilon g{\small (}\phi
 {\small )}=0}\nonumber \\
 &=& \frac{1}{\epsilon }\left( -2\left( -\epsilon g{\small (}\phi
 {\small )}\right) \right)
 =2g{\small (}\phi {\small )}.  \label{39}
 \end{eqnarray}
 Solving for \textbf{$X$}$_{1}^{1}$ in
 \begin{eqnarray}
 \mathbf{X}_{1}E_{0}^{1} &{\LARGE \mid }&_{E_{0}^{1}=0}+H_{1}=0 ,
 \nonumber \\
 \mathbf{X}_{1}E_{0}^{2} &{\LARGE \mid }&_{E_{0}^{2}=0}+H_{2}=0,  \label{40}
 \end{eqnarray}
 gives the same equations as (\ref{14a})-(\ref{21b}%
 ) with a change in (\ref{13a}) and (\ref{13b}%
 ) given by
 \begin{eqnarray}
 \eta _{ss}^{1}+2f{\small (}\theta {\small )} &=&0,  \label{41a} \\
 \eta _{ss}^{2}+2g{\small (}\phi {\small )} &=&0.  \label{41b}
 \end{eqnarray}
 Equations (\ref{18}), (\ref{19}) and (%
 \ref{21a}) yield
 \begin{equation}
 \xi =b_{1}{\small (}s{\small )}.  \label{42}
 \end{equation}
 Substituting the value of $\xi $ into (\ref{16a}) and solving with
 (\ref{41a}), we get
 \begin{equation}
 \eta ^{1}=f_{1}{\small (}\theta ,\phi {\small )}s+f_{2}{\small (}\theta
 ,\phi {\small )}-f{\small (}\theta {\small )}s^{2}.  \label{43}
 \end{equation}
 Integrating (\ref{43}) twice with respect to $\theta $, and using
 (\ref{16a}), we get
 \begin{equation}\nonumber
 \eta _{\theta \theta }^{1}=f_{1\theta \theta }{\small (}\theta ,\phi {\small %
 )}s+f_{2\theta \theta }{\small (}\theta ,\phi {\small )}-f_{\theta \theta }%
 {\small (}\theta {\small )}s^{2}=0.
 \end{equation}
 Comparing the coefficients of the powers of $s$, we obtain
 \begin{equation}
 f_{1\theta \theta }=0, f_{2\theta \theta }=0, f_{\theta
 \theta }=0,  \label{44}
 \end{equation}
 which implies that $f_{1},$ $f_{2}$ and $f$ are linear in $\theta .$ Hence
 we can write
 \begin{eqnarray}
 f_{1}{\small (}\theta ,\phi {\small )} &=&f_{1}{\small (}\phi {\small )}%
 \theta +f_{3}{\small (}\phi {\small )},  \nonumber \\
 f_{2}{\small (}\theta ,\phi {\small )} &=&f_{2}{\small (}\phi {\small )}%
 \theta +f_{4}{\small (}\phi {\small )},  \nonumber \\
 f{\small (}\theta {\small )} &=&c\theta +d,  \label{45}
 \end{eqnarray}
 so that (\ref{43}) becomes
 \begin{equation}
 \eta ^{1}=\left[ f_{1}{\small (}\phi {\small )}s+f_{2}{\small (}\phi {\small %
 )}\right] \theta +f_{3}{\small (}\phi {\small )}s+f_{4}{\small (}\phi
 {\small )}-\left( c\theta +d\right) s^{2}.  \label{46}
 \end{equation}
 From (\ref{41b}), (\ref{14b}) and (\ref%
 {16b}), we obtain
 \begin{equation}
 \eta _{s}^{2}=0,  \label{47}
 \end{equation}
 therefore (\ref{41b}) gives
 \begin{equation}\nonumber
 \eta _{ss}^{2}+2g{\small (}\phi {\small )}=0,
 \end{equation}
 so that
 \begin{equation}
 g{\small (}\phi {\small )}=0,  \label{47a}
 \end{equation}
 \begin{equation}\nonumber
 \eta ^{2}=-\cot \theta f_{6}{\small (}\phi {\small )}+f_{7}{\small (}\phi
 {\small )}.
 \end{equation}
 Substituting the values of $\eta ^{1}$ and $\eta ^{2}$ into (\ref{15a}) we get
 \begin{equation}\nonumber
 \eta _{s\phi }^{1}=0, f_{1}{\small (}\phi {\small )}=A, %
 f_{3}{\small (}\phi {\small )}=B,
 \end{equation}
 which gives
 \begin{equation}
 \eta ^{1}=\left[ As+f_{2}{\small (}\phi {\small )}\right] \theta +Bs+f_{4}%
 {\small (}\phi {\small )}-\left( c\theta +d\right) s^{2}.  \label{48}
 \end{equation}
 Thus (\ref{14a}) yields
 \begin{equation}\nonumber
 b_{1}\left( s\right) =As^{2}-\frac{2}{3}cs^{3}+c_{1}s+c_{0},
 \end{equation}
 so that (\ref{42}) becomes
 \begin{equation}
 \xi =As^{2}-\frac{2}{3}cs^{3}+c_{1}s+c_{0}.  \label{49}
 \end{equation}
 Now, inserting the values of $\eta ^{1}$ and $\eta ^{2}$ in (\ref{17a})
 and (\ref{20a}) gives
 \begin{eqnarray*}
 f_{2}^{\prime }{\small (}\phi {\small )}-\cot \theta f_{6}{\small (}\phi
 {\small )}-\cot \theta f_{2}^{\prime }{\small (}\phi {\small )}\theta -\cot
 \theta f_{4}^{\prime }{\small (}\phi {\small )}=0, \\
 f_{2}^{\prime \prime }{\small (}\phi {\small )}\theta +f_{4}^{\prime \prime }%
 {\small (}\phi {\small )}+\frac{1}{2}sA\sin 2\theta +\frac{1}{2}\sin 2\theta
 f_{2}{\small (}\phi {\small )}-\frac{1}{2}s^{2}c\sin 2\theta \\
 -s\theta A\cos 2\theta -\theta \cos 2\theta f_{2}{\small (}\phi {\small )}%
 -sB\cos 2\theta -\cos 2\theta f_{4}{\small (}\phi {\small )} \\
 +\cos 2\theta \left( c\theta +d\right) s^{2}+2\cos ^{2}\theta f_{6}^{\prime }%
 {\small (}\phi {\small )}-\sin 2\theta f_{7}^{\prime }{\small (}\phi {\small %
 )}=0.
 \end{eqnarray*}
 Comparing the coefficients of the powers of $\theta $ and $s$, we obtain
 \begin{equation}
 c=0, d=0, f{\small (}\theta {\small )}=0.  \label{50a}
 \end{equation}
 From equations (\ref{47a}) and (\ref{50a}) it is clear that, for this case,
 we do not have any new symmetry. Similarly, we see that Cases (iii),
 (iv) and (v) also do not give any non-trivial symmetry.

 \subsection{Approximate symmetries: Second approach}

 Now we adopt another approach and take a more general function to make
 the exact geodesic equations (\ref{36a}) perturbed, and investigate
 the existence of approximate symmetries. Here we approximate the
 geodesic equations of a sphere as
 \begin{eqnarray}
 E_{1}^{1} &:& \ddot{\theta }-\sin \theta \cos \theta \dot{\phi }^{2}+
 \epsilon f{\small (}\theta ,\phi ,\dot{\theta },\dot{\phi }{\small )}=0,  \nonumber \\
 E_{1}^{2} &:& \ddot{\phi }+2\cot \theta \dot{\theta }%
 \dot{\phi }+\epsilon g{\small (}\theta ,\phi ,\dot{\theta },%
 \dot{\phi }{\small )}=0.  \label{51}
 \end{eqnarray}
 As these are second order ordinary differential equations,
 we apply the second prolongation
 \begin{equation}
 \mathbf{X}=\xi \frac{\partial }{\partial s}+\eta ^{1}\frac{\partial }{%
 \partial \theta }+\eta ^{2}\frac{\partial }{\partial \phi }+\eta _{,s}^{1}%
 \frac{\partial }{\partial \dot{\theta }}+\eta _{,s}^{2}\frac{\partial
 }{\partial \dot{\phi }}+\eta _{,ss}^{1}\frac{\partial }{\partial
 \ddot{\theta }}+\eta _{,ss}^{2}\frac{\partial }{\partial \ddot{%
 \phi }},  \label{52}
 \end{equation}
 where
 \begin{equation}
 \xi =\xi _{0}+\epsilon \xi _{1} ,   \label{53}
 \end{equation}
 \begin{equation}
 \eta =\eta _{0}+\epsilon \eta _{1} ,   \label{54}
 \end{equation}
 and the infinitesimal generators, as before, are given by
 \begin{equation}
 \mathbf{X}=\mathbf{X}_{0}+\epsilon \mathbf{X}_{1} .   \label{55}
 \end{equation}
 Equations (\ref{51}) are approximately invariant under the
 approximate group of transformations with the generator given in (\ref%
 {52}) if and only if
 \begin{eqnarray}
 \mathbf{X}E^{1} &{\LARGE \mid }&_{E^{1}=0=E^{2}}=0,  \nonumber \\
 \mathbf{X}E^{2} &{\LARGE \mid }&_{E^{1}=0=E^{2}}=0,  \label{56}
 \end{eqnarray}
 so that we have
 \begin{eqnarray}
 \left[ \mathbf{X}\left( \ddot{\theta }-\sin \theta \cos \theta
 \dot{\phi }^{2}+\epsilon f{\small (}\theta ,\phi ,\dot{\theta }%
 ,\dot{\phi }{\small )}\right) \right] &{\LARGE \mid }&_{\ddot{%
 \theta }-\sin \theta \cos \theta \dot{\phi }^{2}+\epsilon f{\small (}%
 \theta ,\phi ,\dot{\theta },\dot{\phi }{\small )}=0}=0,  \nonumber
 \\
 \left[ \mathbf{X}\left( \ddot{\phi }+2\cot \theta \dot{\theta }%
 \dot{\phi }+\epsilon g{\small (}\theta ,\phi ,\dot{\theta },%
 \dot{\phi }{\small )}\right) \right] &{\LARGE \mid }&_{\ddot{%
 \phi }+2\cot \theta \dot{\theta }\dot{\phi }+\epsilon g{\small %
 (}\theta ,\phi ,\dot{\theta },\dot{\phi }{\small )}=0}=0.
 \label{57}
 \end{eqnarray}
 Now, let us take the following form of the functions \textit{f} and \textit{g}
 introduced in (\ref{51})
 \begin{eqnarray}
 f{\small (}\theta ,\phi ,\dot{\theta },\dot{\phi }{\small )}
 &=&k_{1}+k_{2}\theta +k_{3}\phi +k_{4}\dot{\theta }+k_{5}\dot{%
 \phi }+k_{6}\dot{\theta }^{2}+k_{7}\dot{\phi }^{2},  \nonumber \\
 g{\small (}\theta ,\phi ,\dot{\theta },\dot{\phi }{\small )}
 &=&h_{1}+h_{2}\theta +h_{3}\phi +h_{4}\dot{\theta }+h_{5}\dot{%
 \phi }+h_{6}\dot{\theta }^{2}+h_{7}\dot{\phi }^{2},  \label{58}
 \end{eqnarray}
 where $k_{1},...,k_{7},h_{1},...,h_{7}$ are all constants. Then (\ref%
 {57}) yields
 \begin{eqnarray*}
 \lbrack \eta _{,ss}^{1}-2\sin \theta \cos \theta \dot{\phi }\eta
 _{,s}^{2}-\eta ^{1}\left( \cos ^{2}\theta -\sin ^{2}\theta \right) \dot{\phi }^{2}+\epsilon (k_{2}\eta ^{1}+k_{3}\eta ^{2} \\
 +k_{4}\eta _{,s}^{1}+k_{5}\eta _{,s}^{2}+2k_{6}\dot{\theta }\eta
 _{,s}^{1}+2k_{7}\dot{\phi }\eta _{,s}^{2})]{\LARGE \mid }%
 _{E_{1}=0=E_{2}}=0,
 \end{eqnarray*}
 \begin{eqnarray}
 \lbrack \eta _{,ss}^{2}+2\cot \theta \left( \dot{\phi }\eta _{,s}^{1}+%
 \dot{\theta }\eta _{,s}^{2}\right) -2\dot{\theta }\dot{%
 \phi }\csc ^{2}\theta \eta ^{1}+\epsilon (h_{2}\eta ^{1}+h_{3}\eta ^{2}
 \nonumber \\
 +h_{4}\eta _{,s}^{1}+h_{5}\eta _{,s}^{2}+2h_{6}\dot{\theta }\eta
 _{,s}^{1}+2h_{7}\dot{\phi }\eta _{,s}^{2})]{\LARGE \mid }%
 _{E_{1}=0=E_{2}}=0.  \label{59}
 \end{eqnarray}
 Inserting (\ref{7}) -(\ref{10}) into (\ref{59}), we obtain
 \begin{eqnarray*}
 \lbrack \eta _{ss}^{1}+\dot{\theta }\left( 2\eta _{s\theta }^{1}-\xi
 _{ss}\right) +\dot{\theta }^{2}\left( \eta _{\theta \theta }^{1}-2\xi
 _{s\theta }\right) -\dot{\theta }^{3}\xi _{\theta \theta }+2\dot{\phi }\eta _{s\phi }^{1}+\dot{\phi }^{2}\eta _{\phi \phi }^{1} \\
 +2\dot{\theta }\dot{\phi }\left( \eta _{\theta \phi }^{1}-\xi
 _{s\phi }\right) -2\dot{\theta }^{2}\dot{\phi }\xi _{\theta
 \phi }-\dot{\theta }\dot{\phi }^{2}\xi _{\phi \phi }+\ddot{\theta }\left( \eta _{\theta }^{1}-2\xi _{s}-3\dot{\theta }\xi
 _{\theta }\right) \\
 -2\ddot{\theta }\dot{\phi }\xi _{\phi }+\ddot{\phi }%
 \eta _{\phi }^{1}-\dot{\theta }\ddot{\phi }\xi _{\phi }]-\sin
 2\theta \dot{\phi }[\eta _{s}^{2}+\dot{\phi }\left( \eta
 _{\phi }^{2}-\xi _{s}\right) +\dot{\theta }\eta _{\theta }^{2}-%
 \dot{\phi ^{2}}\xi _{\phi } \\
 -\dot{\theta }\dot{\phi }\xi _{\theta }]-\eta ^{1}\cos 2\theta
 \dot{\phi }^{2}|_{E_{1}=0=E_{2}}=0,
 \end{eqnarray*}
 \begin{eqnarray}
 \lbrack \eta _{ss}^{2}+2\dot{\theta }\eta _{s\theta }^{2}+\dot{%
 \theta }^{2}\eta _{\theta \theta }^{2}+\dot{\phi }\left( 2\eta
 _{s\phi }^{2}-\xi _{ss}\right) +\dot{\phi }^{2}\left( \eta _{\phi
 \phi }^{2}-2\xi _{s\phi }\right) -\dot{\phi }^{3}\xi _{\phi \phi }+2%
 \dot{\theta }\dot{\phi }\left( \eta _{\theta \phi }^{2}-\xi
 _{s\theta }\right)  \nonumber \\
 -\dot{\theta }^{2}\dot{\phi }\xi _{\theta \theta }-2\dot%
 {\theta }\dot{\phi }^{2}\xi _{\theta \phi }+\ddot{\theta }\eta
 _{\theta }^{2}-\ddot{\theta }\dot{\phi }\xi _{\theta }+\ddot{\phi }\left( \eta _{\phi }^{2}-2\xi _{s}-3\dot{\phi }\xi _{\phi
 }\right) -2\dot{\theta }\ddot{\phi }\xi _{\theta }]+2\cot
 \theta  \nonumber \\
 (\dot{\phi }\{\eta _{s}^{1}+\dot{\theta }\left( \eta _{\theta
 }^{1}-\xi _{s}\right) -\dot{\theta }^{2}\xi _{\theta }+\dot{%
 \phi }\eta _{\phi }^{1}-\dot{\theta }\dot{\phi }\xi _{\phi }\}+%
 \dot{\theta }\{\eta _{s}^{2}+\dot{\phi }\left( \eta _{\phi
 }^{2}-\xi _{s}\right)  \nonumber \\
 +\dot{\theta }\eta _{\theta }^{2}-\dot{\phi ^{2}}\xi _{\phi }-%
 \dot{\theta }\dot{\phi }\xi _{\theta }\})-2\dot{\theta }%
 \dot{\phi }\csc ^{2}\theta \eta ^{1}{\LARGE \mid }_{E_{1}=0=E_{2}}=0.
 \label{60}
 \end{eqnarray}
 Inserting the values of $\ddot{\theta }$ and
 $\ddot{\phi }$ from the geodesic equations (\ref{51}), the
 above equations take the form
 \begin{eqnarray*}
 \eta _{ss}^{1}+\dot{\theta }\left( 2\eta _{s\theta }^{1}-\xi
 _{ss}\right) +\dot{\theta }^{2}\left( \eta _{\theta \theta }^{1}-2\xi
 _{s\theta }\right) -\dot{\theta }^{3}\xi _{\theta \theta }+2\dot{\phi }\eta _{s\phi }^{1}+\dot{\phi }^{2}\eta _{\phi \phi }^{1} \\
 +2\dot{\theta }\dot{\phi }\left( \eta _{\theta \phi }^{1}-\xi
 _{s\phi }\right) -2\dot{\theta }^{2}\dot{\phi }\xi _{\theta
 \phi }-\dot{\theta }\dot{\phi }^{2}\xi _{\phi \phi }+(\sin
 \theta \cos \theta \dot{\phi }^{2}-\epsilon f) \\
 (\eta _{\theta }^{1}-2\xi _{s}-3\dot{\theta }\xi _{\theta })-2(\sin
 \theta \cos \theta \dot{\phi }^{2}-\epsilon f)\dot{\phi }\xi
 _{\phi }-(2\cot \theta \dot{\theta }\dot{\phi }+\epsilon
 g)\eta _{\phi }^{1} \\
 +\dot{\theta }(2\cot \theta \dot{\theta }\dot{\phi }%
 +\epsilon g)\xi _{\phi }-\sin 2\theta \dot{\phi }[\eta _{s}^{2}+%
 \dot{\phi }\left( \eta _{\phi }^{2}-\xi _{s}\right) +\dot{%
 \theta }\eta _{\theta }^{2}-\dot{\phi ^{2}}\xi _{\phi }-\dot{%
 \theta }\dot{\phi }\xi _{\theta } \\
 -\eta ^{1}\cos 2\theta \dot{\phi }^{2}+\epsilon (k_{2}\eta
 ^{1}+k_{3}\eta ^{2}+(k_{4}+2k_{6}\dot{\theta })(\eta _{s}^{1}+\dot{\theta }\left( \eta _{\theta }^{1}-\xi _{s}\right) -\dot{\theta }%
 ^{2}\xi _{\theta }+\dot{\phi }\eta _{\phi }^{1} \\
 -\dot{\theta }\dot{\phi }\xi _{\phi })+(k_{5}+2k_{7}\dot%
 {\phi })(\eta _{s}^{2}+\dot{\phi }\left( \eta _{\phi }^{2}-\xi
 _{s}\right) +\dot{\theta }\eta _{\theta }^{2}-\dot{\phi ^{2}}%
 \xi _{\phi }-\dot{\theta }\dot{\phi }\xi _{\theta })=0,
 \end{eqnarray*}
 \begin{eqnarray}
 \eta _{ss}^{2}+2\dot{\theta }\eta _{s\theta }^{2}+\dot{\theta }%
 ^{2}\eta _{\theta \theta }^{2}+\dot{\phi }\left( 2\eta _{s\phi
 }^{2}-\xi _{ss}\right) +\dot{\phi }^{2}\left( \eta _{\phi \phi
 }^{2}-2\xi _{s\phi }\right) -\dot{\phi }^{3}\xi _{\phi \phi }+2%
 \dot{\theta }\dot{\phi }  \nonumber \\
 \left( \eta _{\theta \phi }^{2}-\xi _{s\theta }\right) -\dot{\theta }%
 ^{2}\dot{\phi }\xi _{\theta \theta }-2\dot{\theta }\dot{%
 \phi }^{2}\xi _{\theta \phi }+(\sin \theta \cos \theta \dot{\phi }%
 ^{2}-\epsilon f)\eta _{\theta }^{2}-(\sin \theta \cos \theta \dot{%
 \phi }^{2}  \nonumber \\
 -\epsilon f)\dot{\phi }\xi _{\theta }-(2\cot \theta \dot{%
 \theta }\dot{\phi }+\epsilon g)\left( \eta _{\phi }^{2}-2\xi _{s}-3%
 \dot{\phi }\xi _{\phi }\right) +2\dot{\theta }(2\cot \theta
 \dot{\theta }\dot{\phi }+\epsilon g)\xi _{\theta }+2\cot \theta
 \nonumber \\
 (\dot{\phi }\{\eta _{s}^{1}+\dot{\theta }\left( \eta _{\theta
 }^{1}-\xi _{s}\right) -\dot{\theta }^{2}\xi _{\theta }+\dot{%
 \phi }\eta _{\phi }^{1}-\dot{\theta }\dot{\phi }\xi _{\phi }\}+%
 \dot{\theta }\{\eta _{s}^{2}+\dot{\phi }\left( \eta _{\phi
 }^{2}-\xi _{s}\right) +\dot{\theta }\eta _{\theta }^{2}  \nonumber \\
 -\dot{\phi ^{2}}\xi _{\phi }-\dot{\theta }\dot{\phi }%
 \xi _{\theta }\})-2\dot{\theta }\dot{\phi }\csc ^{2}\theta
 \eta ^{1}+\epsilon (h_{2}\eta ^{1}+h_{3}\eta ^{2}+(h_{4}+2h_{6}\dot{%
 \theta })(\eta _{s}^{1}+\dot{\theta }\left( \eta _{\theta }^{1}-\xi
 _{s}\right)  \nonumber \\
 -\dot{\theta }^{2}\xi _{\theta }+\dot{\phi }\eta _{\phi }^{1}-%
 \dot{\theta }\dot{\phi }\xi _{\phi })+(h_{5}+2h_{7}\dot{%
 \phi })(\eta _{s}^{2}+\dot{\phi }\left( \eta _{\phi }^{2}-\xi
 _{s}\right) +\dot{\theta }\eta _{\theta }^{2}-\dot{\phi ^{2}}%
 \xi _{\phi }-\dot{\theta }\dot{\phi }\xi _{\theta })=0.
 \label{61}
 \end{eqnarray}
 Now, we use (\ref{53}) and (\ref{54}) in (\ref{61}), and separate terms with
 $\epsilon^{0}$ and $\epsilon^{1}$, and ignoring terms with $\epsilon ^{2}$. For
 $\epsilon^{0}$ we get

 \begin{eqnarray*}
 \eta _{0ss}^{1}+\dot{\theta }\left( 2\eta _{s\theta }^{1}-\xi
 _{0ss}\right) +\dot{\theta }^{2}\left( \eta _{0\theta \theta
 }^{1}-2\xi _{s\theta }\right) -\dot{\theta }^{3}\xi _{0\theta \theta
 }+2\dot{\phi }\eta _{0s\phi }^{1}+\dot{\phi }^{2}\eta _{0\phi
 \phi }^{1} \\
 +2\dot{\theta }\dot{\phi }\left( \eta _{0\theta \phi }^{1}-\xi
 _{0s\phi }\right) -2\dot{\theta }^{2}\dot{\phi }\xi _{0\theta
 \phi }-\dot{\theta }\dot{\phi }^{2}\xi _{0\phi \phi }+\sin
 \theta \cos \theta \dot{\phi }^{2}(\eta _{0\theta }^{1} \\
 -2\xi _{s}-3\dot{\theta }\xi _{0\theta })-2\sin \theta \cos \theta
 \dot{\phi }^{3}\xi _{0\phi }-2\cot \theta \dot{\theta }\dot{\phi }\eta _{0\phi }^{1}+2\cot \theta \dot{\theta }^{2}\dot%
 {\phi }\xi _{0\phi } \\
 -\sin 2\theta \dot{\phi }[\eta _{0s}^{2}+\dot{\phi }\left(
 \eta _{0\phi }^{2}-\xi _{0s}\right) +\dot{\theta }\eta _{0\theta
 }^{2}-\dot{\phi ^{2}}\xi _{0\phi }-\dot{\theta }\dot{%
 \phi }\xi _{0\theta }]-\eta _{0}^{1}\cos 2\theta \dot{\phi }^{2}=0,
 \end{eqnarray*}
 \begin{eqnarray}
 \eta _{0ss}^{2}+2\dot{\theta }\eta _{0s\theta }^{2}+\dot{%
 \theta }^{2}\eta _{0\theta \theta }^{2}+\dot{\phi }\left( 2\eta
 _{s\phi }^{2}-\xi _{0ss}\right) +\dot{\phi }^{2}\left( \eta _{0\phi
 \phi }^{2}-2\xi _{s\phi }\right) -\dot{\phi }^{3}\xi _{0\phi \phi }\nonumber \\
 +2\dot{\theta }\dot{\phi }\left( \eta _{0\theta \phi }^{2}-\xi
 _{0s\theta }\right)
 -\dot{\theta }^{2}\dot{\phi }\xi _{0\theta \theta }-2\dot{\theta }\dot{\phi }^{2}\xi _{0\theta \phi }+\sin \theta \cos \theta
 \dot{\phi }^{2}\eta _{0\theta }^{2}-\sin \theta \cos \theta \dot{\phi }^{3}\xi _{0\theta } \nonumber \\
 -2\cot \theta \dot{\theta }\dot{\phi }
 \left( \eta _{0\phi }^{2}-2\xi _{s}-3\dot{\phi }\xi _{0\phi }\right)
 +4\cot \theta \dot{\theta }^{2}\dot{\phi }\xi _{0\theta
 }\nonumber \\
 +2\cot \theta (\dot{\phi }\{\eta _{0s}^{1}+\dot{\theta }%
 \left( \eta _{0\theta }^{1}-\xi _{0s}\right) -\dot{\theta }^{2}\xi
 _{0\theta }
 +\dot{\phi }\eta _{0\phi }^{1}-\dot{\theta }\dot{\phi }%
 \xi _{0\phi }\} \nonumber \\
 +\dot{\theta }\{\eta _{0s}^{2}+\dot{\phi }%
 \left( \eta _{0\phi }^{2}-\xi _{0s}\right) +\dot{\theta }\eta
 _{0\theta }^{2}-\dot{\phi ^{2}}\xi _{0\phi }-\dot{\theta }%
 \dot{\phi }\xi _{0\theta }\})-2\dot{\theta }\dot{\phi }%
 \csc ^{2}\theta \eta ^{1}=0.  \label{62}
 \end{eqnarray}
 Comparing the coefficients of the powers of $\dot{\theta}$ and
 $\dot{\phi}$, we obtain a system of partial differential equations,
 which on solving yields the following solution
 \begin{equation}
 \xi _{0}=c_{1}s+c_{0},  \label{63}
 \end{equation}
 \begin{equation}
 \eta _{0}^{1}=c_{3}\cos \phi +c_{4}\sin \phi ,  \label{64}
 \end{equation}
 \begin{equation}
 \eta _{0}^{2}=\cot \theta \left( c_{4}\cos \phi -c_{3}\sin \phi \right)
 +c_{2}.  \label{65}
 \end{equation}
 We follow the same procedure for $\epsilon^{1}$, compare the coefficients of the powers of
 $\dot{\theta }$ and $\dot{\phi }$,
 and obtain a system of partial differential equations. We denote
 \begin{equation}
 \xi _{0s}=A, \eta _{0}^{1}=B, \eta _{0}^{2}=C, \eta
 _{0\phi }^{1}=D, \eta _{0\phi }^{2}=E, \eta _{0\theta }^{2}=F,
 \label{67}
 \end{equation}
 so that the system of equations become
 \begin{eqnarray}
 (\dot{\theta }\dot{\phi })^{0}:\eta
 _{1ss}^{1}+2(k_{1}+k_{2}\theta +k_{3}\phi )A-(h_{1}+h_{2}\theta +h_{3}\phi
 )D+k_{2}B+k_{3}C=0,  \label{67a} \\
 \eta _{ss}^{2}-(k_{1}+k_{2}\theta +k_{3}\phi )F-(h_{1}+h_{2}\theta
 +h_{3}\phi )(E-2A)+h_{2}B+h_{3}C=0,  \label{67b}
 \end{eqnarray}
 \begin{eqnarray}
 \dot{\theta }:2\eta _{1s\theta }^{1}-\xi
 _{1ss}+k_{4}A-h_{4}D+k_{5}F=0,  \label{68a} \\
 2\eta _{1s\theta }^{2}+2\cot \theta \eta
 _{1s}^{2}-k_{4}F-h_{4}E+h_{4}A+h_{5}F=0,  \label{68b}
 \end{eqnarray}
 \begin{eqnarray}
 \dot{\phi }:2\eta _{1s\phi }^{1}-\sin 2\theta \eta
 _{1s}^{2}+k_{5}A-h_{5}D+k_{4}D+k_{5}E=0,  \label{69a} \\
 2\eta _{1s\phi }^{2}-\xi _{1ss}+2\cot \theta \eta
 _{1s}^{1}-k_{5}F+h_{5}A+h_{4}D=0,  \label{69b}
 \end{eqnarray}
 \begin{eqnarray}
 \dot{\theta }^{2}:\eta _{1\theta \theta }^{1}-2\xi _{1s\theta
 }-h_{6}D=0,  \label{70a} \\
 \eta _{1\theta \theta }^{2}+2\cot \theta \eta _{1\theta
 }^{2}-k_{6}F-h_{6}E=0,  \label{70b}
 \end{eqnarray}
 \begin{eqnarray}
 \dot{\phi }^{2}:\eta _{1\phi \phi }^{1}+\sin \theta \cos \theta \eta
 _{1\theta }^{1}-\eta _{1}^{1}\cos 2\theta -\sin 2\theta \eta _{1\phi
 }^{2}-h_{7}D+2k_{7}E=0,  \label{71a} \\
 \eta _{1\phi \phi }^{2}-2\xi _{1s\phi }+\sin \theta \cos \theta \eta
 _{1\theta }^{2}+2\cot \theta \eta _{1\phi }^{1}-k_{7}F+h_{7}E=0,  \label{71b}
 \end{eqnarray}
 \begin{equation}
 \dot{\theta }^{3}:\xi _{1\theta \theta }=0,  \label{72}
 \end{equation}
 \begin{equation}
 \dot{\phi }^{3}:-\xi _{1\phi \phi }-\sin \theta \cos \theta \xi
 _{1\theta }=0,  \label{73}
 \end{equation}
 \begin{eqnarray}
 \dot{\theta }\dot{\phi }:2\left( \eta _{1\theta \phi }^{1}-\xi
 _{1s\phi }\right) -\sin 2\theta \eta _{1\theta }^{2}-2\cot \theta \eta
 _{1\phi }^{1}+2k_{6}D+2k_{7}F=0,  \label{74a} \\
 2\left( \eta _{1\theta \phi }^{2}-\xi _{1s\theta }\right) +2\cot \theta \eta
 _{1\theta }^{1}-2\csc ^{2}\theta \eta _{1}^{1}+2h_{6}D+2h_{7}F=0,
 \label{74b}
 \end{eqnarray}
 \begin{eqnarray}
 \dot{\theta }\dot{\phi }^{2}:-\xi _{1\phi \phi }-\sin \theta
 \cos \theta \xi _{1\theta }=0,  \label{75a} \\
 -2\xi _{1\theta \phi }+2\cot \theta \xi _{1\phi }=0.  \label{75b}
 \end{eqnarray}
 Equations (\ref{72}), (\ref{73}) and (%
 \ref{75a}) yield
 \begin{equation}
 \xi _{1}=b_{1}{\small (}s{\small )}.  \label{76}
 \end{equation}
 Substituting the value of $\xi $ into (\ref{70a}) and solving with
 (\ref{67a}), we get
 \begin{equation}
 k_{3}=0,  \label{77}
 \end{equation}
 \begin{eqnarray}
 \eta _{1}^{1} &=&-(k_{1}+k_{2}\theta )AS^{2}+\frac{1}{2}(h_{1}+h_{2}\theta
 +h_{3}\phi )Ds^{2}  \nonumber \\
 &&-\frac{1}{2}k_{2}Bs^{2}+(g_{1}{\small (}\phi {\small )}\theta +g_{2}%
 {\small (}\phi {\small )})s+g_{3}{\small (}\phi {\small )}\theta +g_{4}%
 {\small (}\phi {\small )}+\frac{1}{2}h_{6}D^{2},  \label{78}
 \end{eqnarray}
 where $g_{1}${\small (}$\phi ${\small )}$,$ $g_{2}${\small (}$\phi ${\small )%
 }$,$ $g_{3}${\small (}$\phi ${\small )} and $g_{4}${\small (}$\phi ${\small )%
 } are functions of integration. From (\ref{67b}) and (%
 \ref{68b}), we obtain
 \begin{equation}
 k_{1}=k_{2}=k_{3}=h_{1}=h_{2}=h_{3}=0,  \label{78a}
 \end{equation}
 \begin{equation}
 \eta _{1}^{2}=\frac{1}{2}s[-\cot \theta
 (k_{4}F+h_{4}E-h_{4}A-h_{5}F)-k_{4}F_{\theta }-h_{4}E_{\theta
 }+h_{5}F_{\theta }]+f_{3}(\theta ,\phi ),  \label{79}
 \end{equation}
 where $f_{3}(\theta ,\phi )$ is a function of integration. Using the
 above value of $\eta _{1}^{2}$ in (\ref{70b}), we get
 \begin{equation}
 k_{4}=h_{4}=h_{5}=0,  \label{80}
 \end{equation}
 \begin{equation}
 \eta _{1}^{2}=(\theta \cot \theta -\ln \sin \theta )k_{6}D-\frac{1}{2}\theta
 h_{6}D-\cot \theta g_{5}{\small (}\phi {\small )}+g_{6}{\small (}\phi
 {\small )} , \label{80a}
 \end{equation}
 where $g_{5}${\small (}$\phi ${\small )} and $g_{6}${\small (}$\phi ${\small %
 )} are functions of integration. Using the values of $\eta _{1}^{1}$ and $%
 \eta _{1}^{2}$ in (\ref{69a}), gives
 \begin{eqnarray*}
 g_{1}{\small (}\phi {\small )} &=&c_{5}, \\
 g_{2}{\small (}\phi {\small )} &=&-\frac{1}{2}k_{5}(C+A\phi )+c_{6}.
 \end{eqnarray*}
 Now (\ref{69b}) implies
 \begin{eqnarray}
 k_{5} &=&0,  \label{81} \\
 c_{5} &=&0, \\
 \xi &=&c_{7}s+c_{8}.  \label{82}
 \end{eqnarray}
 Similarly (\ref{74a}) and (\ref{71a}) give
 \begin{eqnarray}
 h_{6}=0,  \label{83} \\
 g_{3}{\small (}\phi {\small )}=k_{6}B+c_{9,} \\
 k_{6}=0,  \label{84} \\
 c_{9}=0, \\
 g_{4}{\small (}\phi {\small )}=c_{10}\cos \phi +c_{11}\sin \phi -\frac{\phi
 }{2}\left( k_{7}\tan \theta F+h_{7}B\right) , \\
 g_{5}{\small (}\phi {\small )}=c_{10}\sin \phi -c_{11}\cos \phi -\frac{k_{7}%
 }{2}\tan \theta \csc ^{2}\theta D \nonumber \\
 -\frac{h_{7}}{2}B-\frac{h_{7}}{2}\phi
 D+k_{7}\tan \theta F(\frac{\phi }{2}+1),
 \end{eqnarray}
 \begin{equation}
 g_{4\phi \phi }-\cos 2\theta g_{4}+2\cos ^{2}\theta g_{5\phi }-\sin 2\theta
 g_{6\phi }-h_{7}D+2k_{7}E=0.
 \end{equation}
 Again, using the values of $\eta _{1}^{1}$ and $\eta _{1}^{2}$ in (\ref%
 {74b}) and (\ref{71b}), we get
 \begin{eqnarray}
 g_{6}{\small (}\phi {\small )} &=&c_{12},  \label{85} \\
 h_{7} &=&0, \\
 k_{7} &=&0.  \label{86}
 \end{eqnarray}
 Therefore, we have
 \begin{equation}
 \xi _{1}=c_{7}s+c_{8},
 \end{equation}
 \begin{equation}
 \eta _{1}^{1}=c_{10}\cos \phi +c_{11}\sin \phi ,
 \end{equation}
 \begin{equation}
 \eta _{1}^{2}=\cot \theta \left( c_{11}\cos \phi -c_{10}\sin \phi \right) +c,
 \end{equation}
 which are trivial, as all the constants
 $k_{1},...,k_{7},h_{1},...,h_{7}$ in (\ref{58}) become zero. Hence
 we see that this case also does not give any non-trivial approximate
 symmetries.

 \section{Conclusion}

 One of the methods to solve differential equations is the symmetry method
 by which we can solve differential equations, reduce the order of an
 ordinary differential equation and can reduce the number of independent
 variables in a partial differential equation if it is invariant under a
 one-parameter Lie group of point transformations. The method of `canonical
 variables' is one of the basic procedures for the integration of ordinary
 differential equations with known symmetries. Conservation laws for the
 Euler-Lagrange equations can be constructed when their symmetries are known
 and hence Lagrangians can be constructed.

 For tackling differential equations with a small parameter, the
 method of `approximate symmetries' can be used. Considering the
 geodesic equations of a sphere as a system of two ordinary
 differential equations we have investigated their approximate
 symmetries. Two different approaches have been adopted to find these
 approximate symmetries. Firstly, by converting the geodesic
 equations into perturbed equations by adding the general functions
 $f(\theta )$ and $g(\phi )$ and taking a symmetry of the geodesic
 equations of the sphere. In the second approach we use the
 functions
 \begin{eqnarray*}
 f{\small (}\theta ,\phi ,\dot{\theta },\dot{\phi }{\small )}
 &=&k_{1}+k_{2}\theta +k_{3}\phi +k_{4}\dot{\theta }+k_{5}\dot{%
 \phi }+k_{6}\dot{\theta }^{2}+k_{7}\dot{\phi }^{2}, \\
 g{\small (}\theta ,\phi ,\dot{\theta },\dot{\phi }{\small )}
 &=&h_{1}+h_{2}\theta +h_{3}\phi +h_{4}\dot{\theta }+h_{5}\dot{%
 \phi }+h_{6}\dot{\theta }^{2}+h_{7}\dot{\phi }^{2},
 \end{eqnarray*}
 to convert the geodesic equations into perturbed equations. In both the
 cases, no non-trivial approximate symmetry has been found.

 It would be interesting to see if one can obtain non-trivial approximate symmetries
 for these manifolds by some other method.

\acknowledgments

The authors are thankful to Fazal M. Mahomed for useful discussions.
A research grant from the Higher Education Commission of Pakistan is
gratefully acknowledged.


\begin{thebibliography}{999}


\bibitem{blm} G. W. Bluman, S. Kumei, \textit{Symmetries and differential
equations}, Springer-Verlag, 1989.

\bibitem{ibr} N. H. Ibragimov, \textit{Elementary Lie group analysis and
ordinary differential equations}, John Wiley and Sons, 1999.

\bibitem{ptr} P. E. Hydon, \textit{Symmetry methods for differential
equations: A beginner's guide, }Cambridge University Press, 2000.


\bibitem{C} F. M. Mahomed, C. Qu, \textit{Approximate conditional symmetries
for partial differential equations, }J. Phys. A: Math Gen.
\textbf{33} (2000) 343-356.

\bibitem{D} M. Shih, E. Momoniat, F. M. Mahomed, \textit{Approximate
conditional symmetries and approximate solutions of the perturbed
Fitzhugh-Nagumo equation, }J. Phys. A: Math Gen. \textbf{46} (2005)
023503-(1-10).

\bibitem{A} T. Feroze, F. M. Mahomed, A. Qadir, \textit{The connection
between isometries and symmetries of geodesic equations of the
underlying spaces}, Nonlinear Dynamics, \textbf{45} (2005) 65-74.

\bibitem{oni} B. O'Neil, \textit{Semi Riemannian geometry with applications
to relativity}, Academic Press, 1983.



\bibitem{onl} B. O'Neil, \textit{Elementary differential geometry}, Academic
Press, 1997.

\bibitem{miln} R. S. Milmann, G. D. Parker, \textit{Elements of differential
geometry}, Prentice-Hall, 1977.



\end{thebibliography}
\end{document}